\documentclass[a4paper,11pt]{article}
\usepackage{graphicx,epsfig,picins}
\usepackage{fancyhdr,fancybox}
\usepackage{indentfirst}
\usepackage{titlesec}
\usepackage{verbatim}
\usepackage[sort&compress, numbers]{natbib}
\usepackage{array,dcolumn,tabularx}
\usepackage{setspace}
\usepackage{geometry}
\usepackage{extarrows,chemarrow,xypic} % 画公式中的箭头
\usepackage[small]{caption2}
\usepackage{lineno}
\usepackage{microtype}
\DisableLigatures[f]{encoding = *, family = * }

\usepackage{times}     % 包括 Times Roman + Helvetica + Courier
\usepackage{latexsym}
\usepackage{amsmath}                 % AMS LaTeX 宏包
\usepackage{amssymb}                 % 用来排版漂亮的数学公式
\usepackage{amsbsy}
\usepackage{amsthm}
\usepackage{amsfonts}
\usepackage{mathrsfs}                % 英文花体字体
\usepackage{bm}                      % 数学公式中的黑斜体
\usepackage{arydshln}
\usepackage{relsize}                 % 调整公式字体大小：\mathsmaller, \mathlarger
\usepackage{hyperref}  % 生成书签

\newcommand{\paperfont}{\fontsize{11pt}{1.2\baselineskip}\selectfont}
\begin{document}
%\linenumbers

%%%%%%%%%% 定理类环境的定义 %%%%%%%%%%
\theoremstyle{definition}
\makeatletter
\thm@headfont{\bf}
\makeatother
\newtheorem{definition}{Definition}
\newtheorem{example}{Example}
\newtheorem{theorem}{Theorem}
\newtheorem{lemma}{Lemma}
\newtheorem{corollary}{Corollary}
\newtheorem{remark}{Remark}
\newtheorem{proposition}{Proposition}

%%%%%%%%%% 页眉和页脚的设置 %%%%%%%%%%
\lhead{}
\rhead{}
\lfoot{}
\rfoot{}

%%%%%%%%%% 一些重定义 %%%%%%%%%%
\renewcommand{\refname}{References}
\renewcommand{\figurename}{Fig.}
\renewcommand{\tablename}{Table}
\renewcommand{\proofname}{Proof}

%%%%%%%%%% 符号重定义 %%%%%%%%%%
\newcommand{\diag}{\mathrm{diag}}
\newcommand{\tr}{\mathrm{tr}}
\newcommand{\dnum}{\mathrm{d}}
\newcommand{\Enum}{\mathbb{E}}
\newcommand{\Pnum}{\mathbb{P}}
\newcommand{\Rnum}{\mathbb{R}}
\newcommand{\Cnum}{\mathbb{C}}
\newcommand{\Znum}{\mathbb{Z}}
\newcommand{\Nnum}{\mathbb{N}}
\newcommand{\abs}[1]{\left\vert#1\right\vert}
\newcommand{\set}[1]{\left\{#1\right\}}
\newcommand{\norm}[1]{\left\Vert#1\right\Vert}
\newcommand{\Q}{\boldsymbol{Q}}
\newcommand{\W}{\boldsymbol{W}}
\newcommand{\I}{\boldsymbol{I}}
\newcommand{\M}{\boldsymbol{M}}
\newcommand{\p}{\boldsymbol{p}}
\newcommand{\pai}{\boldsymbol{\pi}}

%%%%%%%%%% 方程按章节编号 %%%%%%%%%%
%\numberwithin{equation}{section}

%%%%%%%%%% 论文标题、作者等 %%%%%%%%%%
\title{\textbf{Kinetic foundation of the zero-inflated negative binomial model for single-cell RNA sequencing data}}
\author{Chen Jia$^{1,2}$ \\
\footnotesize $^1$ Division of Applied and Computational Mathematics, Beijing Computational Science Research Center, Beijing 100193, China. \\
\footnotesize $^2$ Department of Mathematics, Wayne State University, Detroit, MI 48202, U.S.A.\\
\footnotesize Email: chenjia@wayne.edu}
\date{}                              % 日期
\maketitle                           % 生成标题
%\tableofcontents                    % 插入目录
\thispagestyle{empty}                % 首页无页眉页脚

%%%%%%%%%% 正式使用字体 %%%%%%%%%%%
\paperfont

%%%%%%%%%% 摘要 %%%%%%%%%%
\begin{abstract}
Single-cell RNA sequencing data have complex features such as dropout events, over-dispersion, and high-magnitude outliers, resulting in complicated probability distributions of mRNA abundances that are statistically characterized in terms of a zero-inflated negative binomial (ZINB) model. Here we provide a mesoscopic kinetic foundation of the widely used ZINB model based on the biochemical reaction kinetics underlying transcription. Using multiscale modeling and simplification techniques, we show that the ZINB distribution of mRNA abundance and the phenomenon of transcriptional bursting naturally emerge from a three-state stochastic transcription model. We further reveal a nontrivial quantitative relation between dropout events and transcriptional bursting, which provides novel insights into how and to what extent the burst size and burst frequency could reduce the dropout rate. Three different biophysical origins of over-dispersion are also clarified at the single-cell level.
\\

\noindent % 不缩进

%60J27, 60J28, 92C40, 78A70, 92B05
\end{abstract}

%%%%%%%%%% 正文 %%%%%%%%%%
\section{Introduction}
Gene expression in living cells is a complex stochastic process, resulting in spontaneous random fluctuations in mRNA and protein abundances \cite{paulsson2005models}. Recent technological advances in single-cell RNA sequencing (scRNA-seq) have made it possible to measure mRNA expression and provide transcriptome profiles at the single-cell level \cite{sandberg2014entering, eberwine2014promise, kolodziejczyk2015technology, bacher2016design}. Compared with traditional bulk RNA sequencing which measures the average mRNA expression levels across millions of cells, scRNA-seq enables the dissection of gene expression heterogeneity in different cell populations and tissues, and thus allows the investigation of many fundamental biological questions such as the identification of novel cell types, the classification of cell subtypes, and the reconstruction of cellular developmental trajectories \cite{pollen2014low}.

Stochasticity in gene expression measurements has two fundamental origins: (i) the intrinsic noise due to small copy numbers of biochemical molecules and random collisions between them, giving rise to various probabilistic chemical reactions \cite{paulsson2005models}, and (ii) the extrinsic noise due to limitations of current experimental techniques. Although scRNA-seq provides a new level of data resolution, it also produces a much higher noise level than bulk-level measurements. A remarkable characteristic of scRNA-seq data is the high frequency of zero read counts \cite{landau2013dispersion, li2018accurate}. Given the excessive amount of zero observations in scRNA-seq data, it is important to distinguish between (i) the structural (true) zeros where the genes are truly unexpressed and (ii) the dropout (false) zeros where the genes are actually expressed but fail to be detected \cite{liu2016single, hicks2017missing, jia2017accounting, zhu2018unified, gong2018drimpute}. While the former is due to intrinsic biological variability, the latter, which is referred to as dropout events, is due to extrinsic technical reasons.

Due to the tiny amount of mRNA in an individual cell, the input material needs to be captured with low efficiency and go through many rounds of amplification before being sequenced. This results in low mRNA capture rate and strong amplification bias, as well as dropout events \cite{wang2015advances}. To be more specific, the workflow of scRNA-seq experiments includes the following steps: isolation of single cells, cell lysis while preserving mRNA, mRNA capture, reverse transcription of primed RNA into cDNA, cDNA amplification, library preparation, and sequencing \cite{haque2017practical}. During these steps, possible technical reasons leading to dropouts include mRNA degradation after cell lysis, low efficiency of mRNA capture, reverse transcription, and cDNA amplification, library size differences, and sequencing depth \cite{hicks2017missing}. Recent studies \cite{haque2017practical} have shown that the efficiency for poly-adenylated mRNA species to be captured, converted into cDNA, and amplified can range between 10\% and 40\%, depending on the study. This means that if the starting transcripts in an individual cell are in low amount, there is a certain probability that they will not be detected by current scRNA-seq methods.

Besides the dropout effect, other characteristics of scRNA-seq data include over-dispersion \cite{mccarthy2012differential} and high-magnitude outliers \cite{kharchenko2014bayesian} due to the stochastic nature of gene expression at the single-cell level and the related phenomenon of transcriptional bursting \cite{haque2017practical}. Given these complex features of scRNA-seq data, recent studies have highlighted the need to develop novel statistical and computational methods in data analysis, especially differential expression analysis \cite{bacher2016design}. When handling dropout events, a popular perspective held by the bioinformatic field is that the complicated probability distributions of mRNA abundances in a cell population need to be explicitly characterized by a global zero-inflation parameter, resulting in various zero-inflated models \cite{mcdavid2012data, paulson2013differential}. Among these statistical models, the zero-inflated negative binomial (ZINB) model is the most widely used \cite{pierson2015zifa, wagner2016revealing, fang2016zero, vallejos2017normalizing, gao2017nanogrid, wallrapp2017neuropeptide, risso2018general, chen2018umi, lopez2018deep, van2018observation, miao2018desingle, eraslan2019single}, where the zero-inflated part describes dropouts and the negative binomial part accounts for over-dispersion. Some other commonly used models are listed in Sec. \ref{discussion}.

Modern sciences emphasize quantitative characterization of experimental observations, which is widely known as mathematical modeling. Along this line, two types of modeling methods should be distinguished: data-driven and mechanism-based modeling \cite{qian2013stochastic}. The former explains experimental phenomena in terms of data analysis based on various mathematical formulas and statistical models, while the latter understands the world in terms of mathematical deductions based on various mechanisms and scientific laws. The ZINB model of scRNA-seq data proposed in previous studies belongs to the former category.

In the present work, we provide a mesoscopic kinetic foundation of the widely used ZINB model based on the stochastic biochemical reaction kinetics underlying transcription. In fact, many stochastic models of transcription dynamics have been proposed \cite{peccoud1995markovian, raj2006stochastic, iyer2009stochasticity, mugler2009spectral, chong2014mechanism, kumar2015transcriptional, jia2017emergent, klindziuk2018theoretical}. Although some previous models could provide a clear explanation of over-dispersion, very few of them have incorporated the dropout effect into their model assumptions. So far, there is still a lack of a kinetic basis for the ZINB distribution of mRNA abundance. In addition, it is widely believed that the complex features of scRNA-seq data are closely related to the phenomenon of transcriptional bursting. However, the quantitative relationship among dropout events, over-dispersion, and transcriptional bursting still remains unclear. The present paper addresses these issues.

\section{A novel three-state model of transcription}
\begin{figure}[!htb]
\centerline{\includegraphics[width=0.6\textwidth]{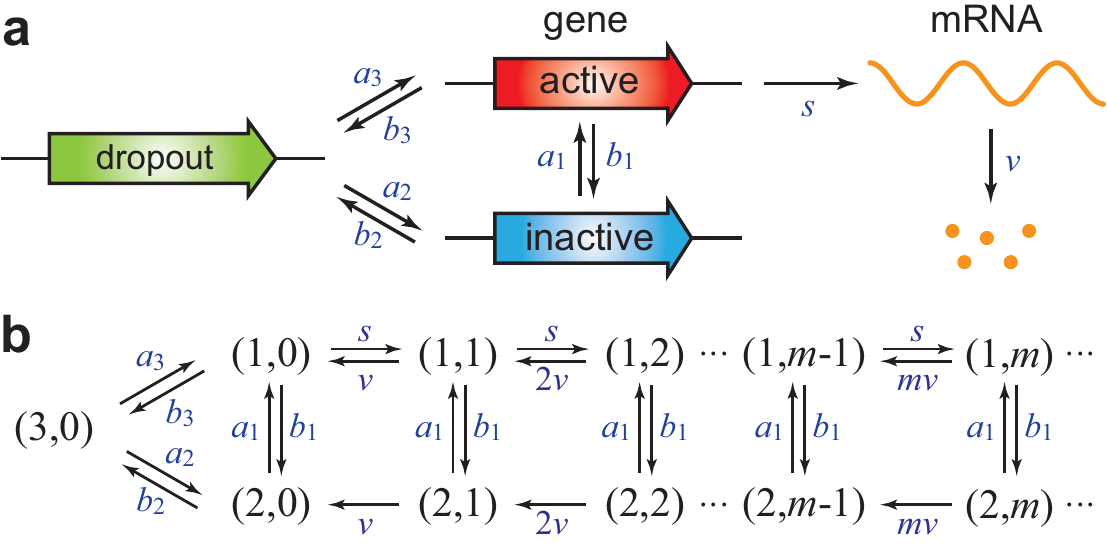}}
\caption{\textbf{Stochastic transcription kinetics in individual cells with dropout events.} (a) A three-state transcription model involving gene switching among the active, inactive, and dropout states. Here the dropout state characterizes the detection state where the mRNA expression of this gene is undetectable. (b) Transition diagram of the Markovian model whose dynamics is governed by the chemical master equation.}\label{model}
\end{figure}

Based on the central dogma of molecular biology, the transcription of a gene in an individual cell has a standard two-stage representation involving the switching of the gene between an active and an inactive epigenetic state and the synthesis of the mRNA from the gene \cite{paulsson2005models}. In the active state, the gene produces the mRNA. When the gene is inactive, the process of mRNA synthesis is terminated. Due to various technical factors in scRNA-seq experiments such as low mRNA capture rate, amplification bias, and sequencing depth, at a particular time, the mRNA expression in a single cell can be either detectable or undetectable \cite{eraslan2019single}. As a result, it is reasonable to assume that the gene of interest can exist in a third state, referred to as the dropout state, where the mRNA expression of this gene cannot be detected due to technical reasons. Here the dropout state should not be regarded as an epigenetic conformation of the gene. Instead, it characterizes an undetectable state where the transcriptional signal of the gene is missing. These considerations lead to the three-state transcription model illustrated in Fig. \ref{model}(a), where a transcript can be synthesized with rate $s$ or be degraded with rate $v$, and the gene can switch among the active, inactive, and dropout states with certain switching rates $a_i$ and $b_i$, $i = 1,2,3$. Compared with the classical two-state transcriptional model without the dropout state \cite{paulsson2005models}, the cyclic structure of gene state switching will remarkably increase the theoretical complexity, as we shall see.

From the chemical perspective, the microstate of the gene of interest can be described by an ordered pair $(i,m)$: the state $i$ of the gene and the copy number $m$ of detectable transcripts, where $i = 1,2,3$ correspond to the active, inactive, and dropout states, respectively. Then the stochastic dynamics of our three-state transcription model can be described by the Markov jump process (continuous-time Markov chain) with transition diagram illustrated in Fig. \ref{model}(b). Since the transcriptional signal is missing when a dropout occurs, it is reasonable to assume that the dropout state can only exist with zero detectable transcript, described by the microstate $(3,0)$.

Experimentally, it was widely observed that the dropout rate for a given cell strongly depends on its expression level, with dropouts being more frequent for cells with low mRNA expression levels \cite{kharchenko2014bayesian}. In general, the total content of mRNA in a single cell is low (0.01-2.5pg per cell) \cite{livesey2003strategies} and most genes only transcribe a small copy number of mRNA \cite{taniguchi2010quantifying}. Due to the tiny amount of mRNA in an individual cell, the input material needs to be captured with low efficiency and go through many rounds of amplification before being sequenced. This results in low mRNA capture rate and strong amplification bias, as well as dropout events \cite{risso2018general}. As a result, microstates $(1,m)$ and $(2,m)$ with tiny mRNA abudance $m$ are more likely to convert to the dropout microstate $(3,0)$. In our Markovian model, for simplicity, we assume that $(3,0)$ can only be reached from $(1,m)$ and $(2,m)$ with $m = 0$ (Fig. \ref{model}(b)). In Sec. \ref{discussion}, a removal of this assumption will be discussed and a more realistic model will be given.

There is another reason leading us to consider the three-state transcription model. Recent single-cell experiments have provided evidence that for many genes, more than two states may participate in the transcription process \cite{suter2011mammalian, harper2011dynamic, rieckh2014noise, corrigan2016continuum, bintu2016dynamics}. In fact, if a gene can only switch between the active and inactive states, then the sojourn times in the active and inactive states should be exponentially distributed. However, recent single-cell time-lapse measurements in eukaryotic cells \cite{suter2011mammalian, harper2011dynamic} have indicated that the sojourn time in the inactive state may have a non-exponential peaked distribution. This indicates that the gene dynamics in the inactive state may contain at least two exponential steps, so that in sum the gene would undergo a three-state switching process.

In particular, in two recent studies, the authors monitored gene expression dynamics in mouse fibroblasts \cite{suter2011mammalian} and Chinese hamster ovary cells \cite{bintu2016dynamics} using single-cell time-lapse microscopy and found that both data sets were well described by a three-state gene expression model involving gene switching among an active, an inactive (reversibly silent), and a refractory (irreversibly silent) state. The difference between the inactive and refractory states is that the former has a good chance to switch back to the active state, while the possibility for the latter to switch back is much lower. In the inactive or refractory state, RNA polymerases could either be absent from the promoter or present in a paused state. Therefore, the dropout state in our three-state transcription model may have two different interpretations: It may either correspond to an undetectable state due to purely technical factors or correspond to a refractory state due to real biological factors.

%Recently, the inactive and refractory states have been endowed with some more detailed biological implications. Some authors \cite{suter2011mammalian} described the refractory state as the state where the RNA polymerase is absent from the promoter and described the inactive state as the state where the RNA polymerase is present in a paused state. In \cite{bartman2019transcriptional}, the authors described the refractory state as the state where both the transcription factor and RNA polymerase are absent from the promoter and described the inactive state as the state where the transcription factor bind to the promoter while RNA polymerase does not bind to the promoter. Some other authors \cite{cao2019multi} described the refractory (inactive) state as the state where RNA polymerase are absent from the promoter before (after) chromosome remodeling.

Let $p_{i,m}(t)$ denote the probability of having $m$ detectable transcripts at time $t$ when the gene is in state $i$. Then the evolution of the Markovian model is governed by the chemical master equation
\begin{equation*}\left\{
\begin{split}
\dot p_{1,0} =&\; a_1p_{2,0}+a_3p_{3,0}+vp_{1,1}-(b_1+b_3+s)p_{1,0}, \\
\dot p_{2,0} =&\; b_1p_{1,0}+a_2p_{3,0}+vp_{2,1}-(a_1+b_2)p_{2,0}, \\
\dot p_{3,0} =&\; b_3p_{1,0}+b_2p_{2,0}-(a_2+a_3)p_{3,0}, \\
\dot p_{1,m} =&\; a_1p_{2,m}+sp_{1,m-1}+(m+1)vp_{1,m+1}-(b_1+s+mv)p_{1,m},\;\;\;m\geq 1, \\
\dot p_{2,m} =&\; b_1p_{1,m}+(m+1)vp_{2,m+1}-(a_1+mv)p_{2,m},\;\;\;m\geq 1.
\end{split}\right.
\end{equation*}
Here $s$ is the transcription rate; $v$ is the degradation rate of the mRNA; $a_i$ and $b_i$, $i = 1,2,3$ are the switching rates of the gene among the three states. Since $(i,m)$ represents the microstate of having $m$ transcripts in a single cell when the gene is in state $i$ and each transcript can be degraded with rate $v$, the transition rate from microstate $(i,m)$ to microstate $(i,m-1)$, which represents the total degradation rate of the $m$ transcripts, should be $mv$ (Fig. \ref{model}(b)) \cite{paulsson2005models}. In addition, since the dropout state could describe a refractory state, which has a lower chance to switch back to the active state than the inactive state, it is natural to assume $a_1 > a_3$ in our model.

\section{Model simplification via decimation}
\begin{figure}[!htb]
\centerline{\includegraphics[width=0.7\textwidth]{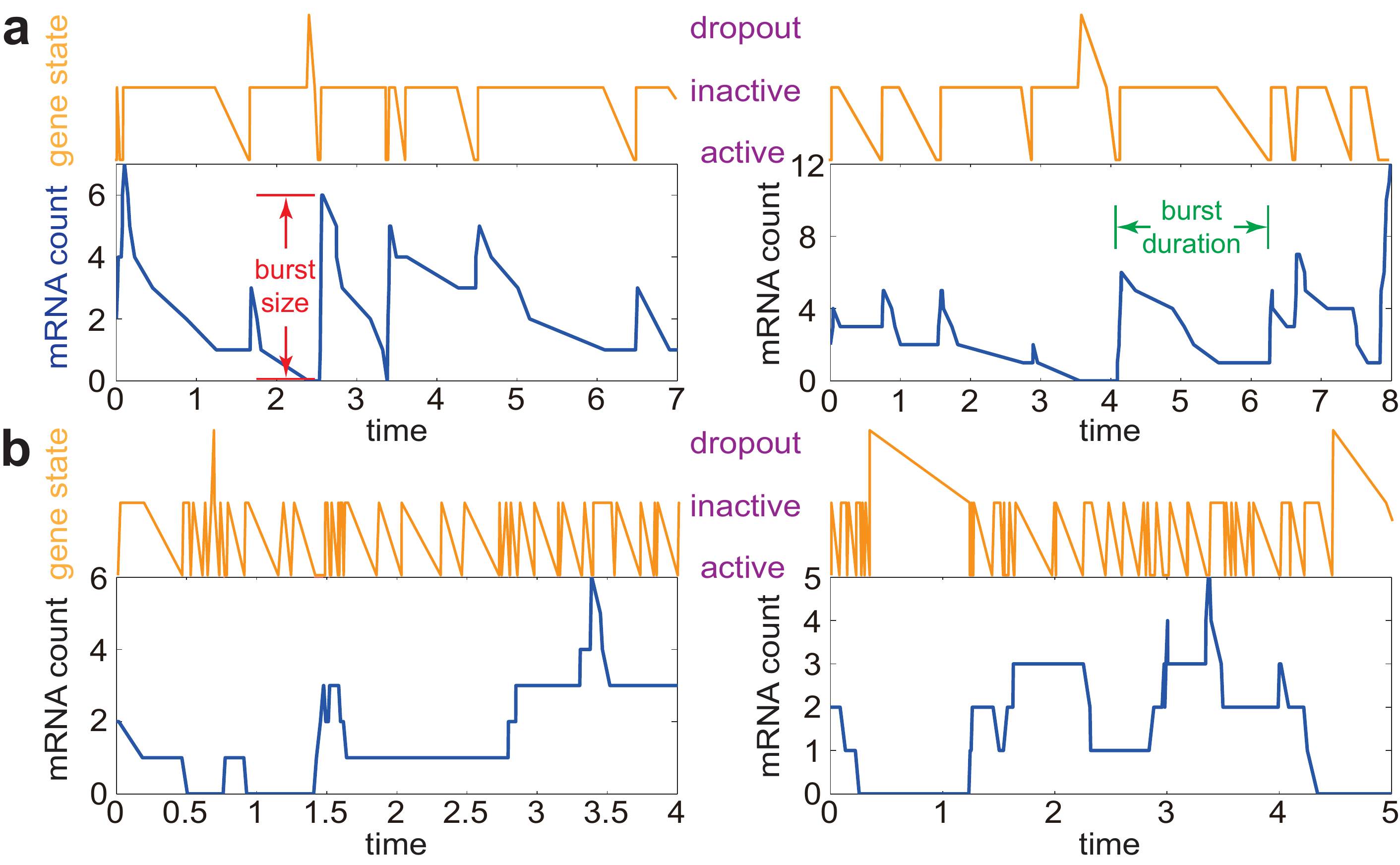}}
\caption{\textbf{Numerical simulations of the stochastic trajectories of the gene state and mRNA copy number based on the original Markovian model under two sets of biologically relevant parameters.}
(a) Two typical trajectories when the mean burst size $h$ and maximum burst frequency $\lambda$ are moderate. The model parameters are chosen as $h = 3, \lambda = 1.5, s = hb_1, v = 1, a_1 = \lambda v, b_1 = 25, a_2 = 1, b_2 = 4, a_3 = 0, b_3 = 1$.
(b) Two typical trajectories in the limiting case of $h\rightarrow 0$ and $\lambda\rightarrow\infty$, while $\lambda h = \gamma$ is kept as a constant. The model parameters are chosen as $h = 0.2, \lambda = 10, s = hb_1, v = 1, a_1 = \lambda v, b_1 = 100, a_2 = 1, b_2 = 4, a_3 = 0, b_3 = 1$.}\label{trajectory}
\end{figure}

One of the most important reasons for over-dispersion of bulk and single-cell RNA-seq data is transcriptional bursting, also known as transcriptional pulsing \cite{haque2017practical}, which describes the phenomenon of relatively short transcriptionally active and high expression periods followed by longer transcriptionally silent and low expression periods \cite{cai2008frequency, suter2011mammalian}, resulting in spontaneous temporal fluctuations of transcript levels (Fig. \ref{trajectory}(a)).

In general, transcriptional bursting results from multiple time scales underlying the transcription process \cite{moran2012sizing}. In fact, the mechanism of transcriptional bursting has been described by Paulsson in his review paper \cite{paulsson2005models}, ``If genes are mostly inactive but transcribe a large number of mRNAs while in the active state, transcription could occur in bursts". Intuitively, if we require the gene to be mostly inactive, the switching rate $b_1$ of the gene from the active to the inactive state should be much larger than the reverse switching rate $a_1$ from the inactive to the active state. On the other hand, if we require the gene to transcribe a large number of transcripts while in the active state, the transcription rate $s$ should be very large, at least at the same order of magnitude as the switching rate $b_1$. These considerations lead to the following biochemical conditions for transcriptional bursting: $b_1\gg a_1$ and $s/b_1$ is finite. Here, by saying that $s/b_1$ is finite, we mean that $s$ and $b_1$ are roughly at the same order of magnitude. When the gene is active, the large transcription rate $s$ will give rise to fast accumulation of mRNA. Once the gene becomes inactive, the transcription process is terminated and transcripts will be degraded until the gene becomes active again. We stress here that the above biochemical conditions imposed on the rate constants are consistent with a recent single-cell experiment on transcriptional bursting \cite{suter2011mammalian}, where the authors monitored the transcription kinetics in mouse fibroblasts using time-lapse bioluminescence imaging and found that the three rate constants $a_1$, $b_1$, and $s$ across different genes are typically at the magnitude of 0.01/min, 0.1/min, and 1/min, respectively.

\begin{figure}[!htb]
\centerline{\includegraphics[width=0.9\textwidth]{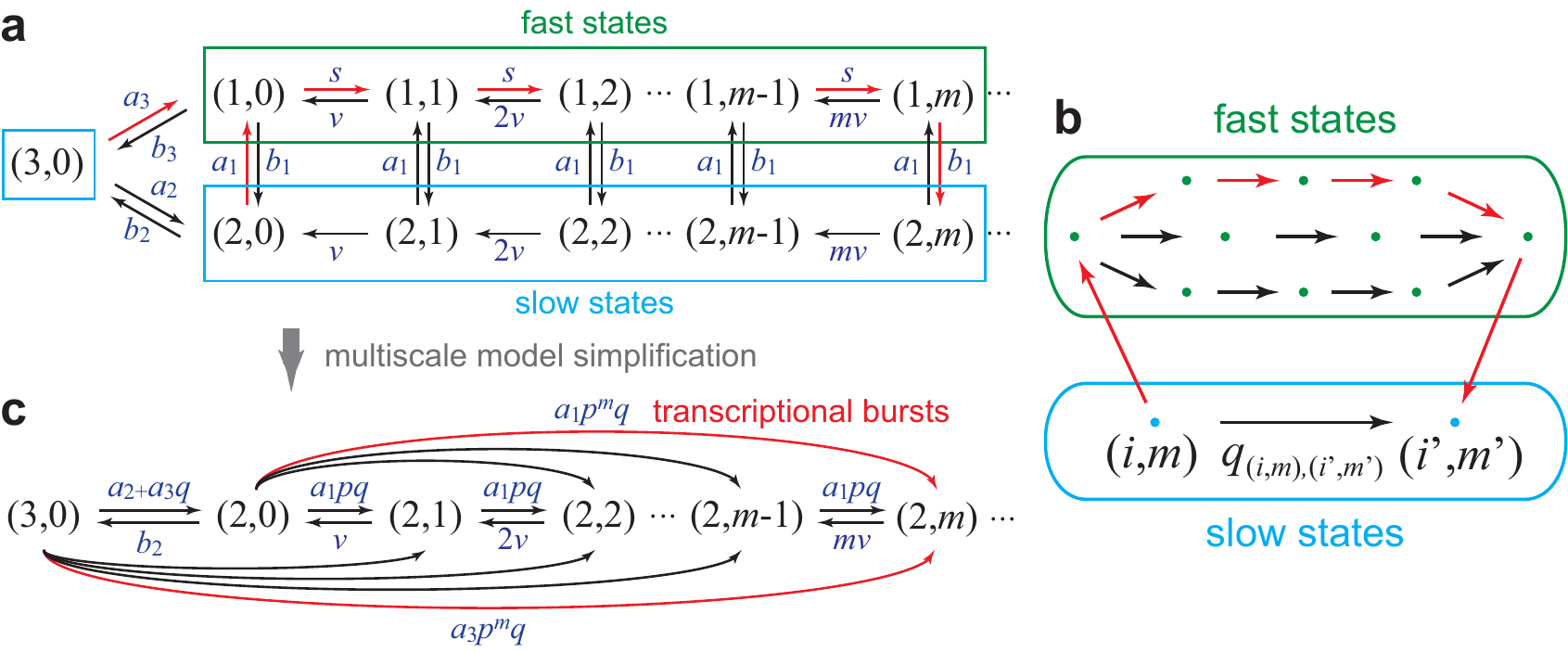}}
\caption{\textbf{Multiscale model simplification of the Markovian model.} (a) Fast (green) and slow (blue) states of the original Markovian model. (b) Schematic diagram of the decimation method of model simplification. The effective transition rate from microstate $(i,m)$ to microstate $(i',m')$ is the superposition of the direct transition rate and the contribution of indirect transitions via all fast transition paths. (c) Transition diagram of the reduced Markovian model when $b_1\gg a_1$ and $s/b_1$ is finite. The red arrows in (a)-(c) point the directions of typical fast transition paths, which correspond to random transcriptional bursts.}\label{decimation}
\end{figure}

Due to the timescale separation of the underlying biochemical reaction kinetics, our Markovian model can be simplified to a much simpler one. To see this, let $\beta = b_1/a_1\gg 1$ denote the ratio of the switching rates between the active and inactive states. Moreover, let $q_{(i,m),(i',m')}$ denote the transition rate of the Markovian model from microstate $(i,m)$ to microstate $(i',m')$ and let
\begin{equation*}
q_{(i,m)} = \sum_{(i',m')\neq(i,m)}q_{(i,m),(i',m')}
\end{equation*}
denote the rate at which the system leaves microstate $(i,m)$, which is defined as the sum of transition rates from $(i,m)$ to other microstates. Since $\beta\gg 1$, we say that $(i,m)$ is a fast state if
\begin{equation*}
\lim_{\beta\rightarrow\infty}q_{(i,m)} = \infty
\end{equation*}
and we say that $(i,m)$ is a slow state if
\begin{equation*}
\lim_{\beta\rightarrow\infty}q_{(i,m)} < \infty.
\end{equation*}
If $(i,m)$ is a fast state, then the time that the system stays in this state will be very short. Since $b_1\gg a_1$ and $s/b_1$ is finite, we write $b_1 = \beta a_1$ and $s = \beta a_1(s/b_1)$, where $\beta\gg 1$ and we treat $a_1$ and $s/b_1$ as constants. Here $b_1$ and $s$ are the only two model parameters depending on $\beta$ and all other model parameters are independent of $\beta$. It is easy to check that the leaving rates of all microstates are given by
\begin{equation*}\left\{
\begin{split}
q_{(1,0)} &= b_1+b_3+s = \beta a_1(1+s/b_1)+b_3, \\
q_{(2,0)} &= a_1+b_2, \\
q_{(3,0)} &= a_2+a_3, \\
q_{(1,m)} &= b_1+s+mv = \beta a_1(1+s/b_1)+mv,\;\;\;m\geq 1, \\
q_{(2,m)} &= a_1+mv,\;\;\;m\geq 1,
\end{split}\right.
\end{equation*}
which shows that
\begin{equation*}
\lim_{\beta\rightarrow\infty}q_{(1,m)} = \infty,\;\;\;
\lim_{\beta\rightarrow\infty}q_{(2,m)} < \infty,\;\;\;
\lim_{\beta\rightarrow\infty}q_{(3,m)} < \infty.
\end{equation*}
Therefore, all active microstates $(1,m)$ are fast states and all other microstates $(2,m)$ and $(3,m)$ are slow states (Fig. \ref{decimation}(a)). By using a classical simplification method of two-time-scale Markov jump processes called decimation \cite{pigolotti2008coarse, cappelletti2016elimination, jia2016reduction, jia2016simplification, bo2016multiple, jia2017simplification}, the original Markovian model can be simplified to a reduced one by removal of all fast states.

The remaining question is to determine the transition diagram and effective transition rates of the reduced model. This process is described as follows. Suppose that the original model jumps from microstate $(i,m)$ to another microstate a particular time. When $\beta\gg 1$, the transition probability from microstate $(i,m)$ to another microstate $(i',m')$ is given by
\begin{equation*}
w_{(i,m),(i',m')} = \lim_{\beta\rightarrow\infty}\frac{q_{(i,m),(i',m')}}{q_{(i,m)}}.
\end{equation*}
Let $(i_1,m_1),\cdots,(i_n,m_n)$ be a sequence of microstates. We say that
\begin{equation*}
c: (i,m)\rightarrow(i_1,m_1)\rightarrow\cdots\rightarrow(i_n,m_n)\rightarrow(i',m')
\end{equation*}
is a fast transition path from $(i,m)$ to $(i',m')$ if the intermediate states $(i_1,m_1),\cdots,(i_n,m_n)$ are all fast states. Moreover, the probability weight $w_c$ of the fast transition path $c$ is defined as
\begin{equation*}
w_c = q_{(i,m),(i_1,m_1)}w_{(i_1,m_1),(i_2,m_2)}\cdots w_{(i_n,m_n),(i',m')}.
\end{equation*}
According to the decimation theory \cite{pigolotti2008coarse, cappelletti2016elimination, jia2016reduction, jia2016simplification, bo2016multiple, jia2017simplification}, the effective transition rate from $(i,m)$ to $(i',m')$ is given by
\begin{equation*}
\tilde{q}_{(i,m),(i',m')} = q_{(i,m),(i',m')}+\sum_cw_c,
\end{equation*}
where $c$ ranges over all fast transition paths from $(i,m)$ to $(i',m')$. This formula shows that the effective transition rate from $(i,m)$ to $(i',m')$ is the sum of two parts: the direct transition rate $q_{(i,m),(i',m')}$ and the contribution of indirect transitions via all fast transition paths, as illustrated in Fig. \ref{decimation}(b).

Since the intermediate states of a fast transition path $c$ are all fast states, in order for the path to have a positive probability weight, all the intermediate transitions along this path must satisfy
\begin{equation*}
\lim_{\beta\rightarrow\infty}q_{(i_1,m_1),(i_2,m_2)} = \cdots = \lim_{\beta\rightarrow\infty}q_{(i_n,m_n),(i',m')} = \infty.
\end{equation*}
By using this criterion, it is easy to see that the original model only has two types of fast transition paths with positive probability weights, which are given by
\begin{equation}\label{path1}
(2,m)\rightarrow(1,m)\rightarrow(1,m+1)\rightarrow\cdots\rightarrow(1,m')\rightarrow(2,m'),\;\;\;m'>m,
\end{equation}
and
\begin{equation}\label{path2}
(3,0)\rightarrow(1,0)\rightarrow(1,1)\rightarrow\cdots\rightarrow(1,m)\rightarrow(2,m),\;\;\;m\geq 0,
\end{equation}
as illustrated by the red arrows in Fig. \ref{decimation}(a). To proceed, we defined two constants $p$ and $q$ as
\begin{equation*}
p = \frac{s}{s+b_1},\;\;\;q = \frac{b_1}{s+b_1}.
\end{equation*}
When $\beta\gg 1$, the transition probabilities along the above two fast transition paths are given by
\begin{gather*}
w_{(1,m),(1,m+1)} = \lim_{\beta\rightarrow\infty}\frac{s}{q_{(1,m)}} = p, \\
w_{(1,m),(2,m)} = \lim_{\beta\rightarrow\infty}\frac{b_1}{q_{(1,m)}} = q.
\end{gather*}
Therefore, the probability weight of the path \eqref{path1} is given by $a_1p^{m'-m}q$ and the probability weight of the path \eqref{path2} is given by $a_3p^mq$. Since there is no direct transition, the effective transition rate from $(2,m)$ to $(2,m')$ is the indirect transition rate via the fast transition path \eqref{path1}:
\begin{equation*}
\tilde{q}_{(2,m),(2,m')} = a_1p^{m'-m}q.
\end{equation*}
Moreover, the effective transition rate from $(3,0)$ to $(2,m)$ is the sum of the direct transition rate and the indirect transition rate via the fast transition path \eqref{path2}:
\begin{equation*}
\begin{split}
\tilde{q}_{(3,0),(2,m)} &= q_{(3,0),(2,m)}+a_3p^mq = \begin{cases}
a_2+a_3q, &m = 0,\\
a_3p^mq, &m\geq 1.
\end{cases}\end{split}
\end{equation*}
The above two formulas indicate that the reduce model may produce large jumps of mRNA abundance within a very short period, which correspond to transcriptional bursts. Each random burst corresponds to a fast transition path of the original model (see the red arrows in Fig. \ref{decimation}(a)). So far, we have obtained all effective transition rates of the reduced model, whose transition diagram is depicted in Fig. \ref{decimation}(c).

The above calculations can be understood intuitively as follows. Since $b_1\gg a_1$ and $s/b_1$ is finite, the process of mRNA synthesis followed by gene silencing is essentially instantaneous. Once the gene becomes active, it can either produce a transcript with probability $p = s/(s+b_1)$ or switch to the inactive state with probability $q = 1-p = b_1/(s+b_1)$. Therefore, the probability that the gene produces $k$ transcripts in a single burst before it is finally silenced will be $p^kq$, which follows a geometric distribution. This consideration again leads to the reduced model illustrated in Fig. \ref{decimation}(c). The evolution of the reduced model is governed by the chemical master equation
\begin{equation}\label{reducedcme}\left\{
\begin{split}
\dot p_{2,0} =&\; vp_{2,1}+(a_2+a_3q)p_{3,0}-(a_1p+b_2)p_{2,0}, \\
\dot p_{3,0} =&\; b_2p_{2,0}-(a_2+a_3)p_{3,0}, \\
\dot p_{2,m} =&\; \sum_{k=0}^{m-1}a_1p^{m-k}qp_{2,k}+(m+1)vp_{2,m+1}\\
&\; +a_3p^mqp_{3,0}-(a_1p+mv)p_{2,m},\;\;m\geq 1.
\end{split}\right.
\end{equation}
Since the burst size of the mRNA, which is defined as the number of transcripts produced in a single burst (Fig. \ref{trajectory}(a)), is geometrically distributed, its expected value is given by
\begin{equation*}
h = \sum_{k=0}^\infty kp^kq = \frac{p}{q} = \frac{s}{b_1}.
\end{equation*}

\section{Theoretical foundation for the ZINB model}
Although the topological structure of the reduced model is complicated, its steady-state probability distribution can be solved analytically. To see this, let $p^{ss}_{(i,m)}$ denote the steady-state probability of microstate $(i,m)$. At the steady state, the probabilities of all microstates are time-independent and thus the left side of \eqref{reducedcme} must equal zero, giving rise to a set of linear equations. Interestingly, this set of linear equations can be solved explicitly with its solution being given by (see Appendix)
\begin{equation}\label{distribution}\left\{
\begin{split}
p^{ss}_{2,0} &= A\cdot\frac{a_1}{\tilde a_1}, \\
p^{ss}_{3,0} &= A\cdot\frac{a_1b_2}{\tilde a_1(a_2+a_3)}, \\
p^{ss}_{2,m} &= A\cdot\frac{p^m(a_1/v)_m}{m!},\;\;\;m\geq 1,
\end{split}\right.
\end{equation}
where $A$ is a normalization constant, $\tilde a_1$ is a constant given by
\begin{equation}\label{a1}
\tilde a_1 = a_1+\frac{b_2a_3}{a_2+a_3},
\end{equation}
and $(x)_m = x(x+1)\cdots(x+m-1)$ is the Pochhammer symbol. Since all steady-state probabilities add up to one, the normalization constant $A$ can be determined as
\begin{equation*}
A = \left[\frac{a_1}{\tilde a_1}\left(1+\frac{b_2}{a_2+a_3}\right)+q^{-a_1/v}-1\right]^{-1}.
\end{equation*}
Let $p^{ss}_m$ denote the steady-state probability of having $m$ copies of detectable transcripts. Then we obtain
\begin{equation*}\left\{
\begin{split}
p^{ss}_0 &= p^{ss}_{2,0}+p^{ss}_{3,0} = A\cdot\frac{a_1}{\tilde a_1}\left(1+\frac{b_2}{a_2+a_3}\right), \\
p^{ss}_m &= p^{ss}_{2,m} = A\cdot\frac{p^m(a_1/v)_m}{m!},\;\;\;m\geq 1.
\end{split}\right.
\end{equation*}
Since the probabilities $p$ and $q$ can be represented by the mean burst size $h$ as
\begin{equation*}
p = \frac{h}{1+h},\;\;\;q = \frac{1}{1+h},
\end{equation*}
the steady-state distribution of mRNA abundance can be written in a unified way as
\begin{equation*}
\begin{split}
p^{ss}_m &= w\delta_0(m)+(1-w)\frac{(\lambda)_m}{m!}
\left(\frac{h}{1+h}\right)^m\left(\frac{1}{1+h}\right)^{\lambda} \\
&= wp^{\textrm{zero-inflated}}_m+(1-w)p^{\textrm{NB}}_m,
\end{split}
\end{equation*}
where $\delta_0(m)$ is Kronecker's delta function which takes the value of 1 when $m = 0$ and takes the value of 0 otherwise, and $\lambda>0$ and $0<w<1$ are two constants given by
\begin{gather*}
\lambda = \frac{a_1}{v},\\
w = \frac{\frac{a_1}{\tilde a_1}\left(1+\frac{b_2}{a_2+a_3}\right)-1}
{\frac{a_1}{\tilde a_1}\left(1+\frac{b_2}{a_2+a_3}\right)+(1+h)^{\lambda}-1}.
\end{gather*}
Here $0<w<1$ is a result of our model assumption $a_1> a_3$. This is exactly the ZINB distribution of mRNA abundance widely used in scRNA-seq data analysis \cite{pierson2015zifa, wagner2016revealing, fang2016zero, vallejos2017normalizing, gao2017nanogrid, wallrapp2017neuropeptide, risso2018general, chen2018umi, lopez2018deep, van2018observation, miao2018desingle, eraslan2019single}. Specifically, the ZINB distribution is the mixture of two distributions: the zero-inflated part
\begin{equation*}
p^{\textrm{zero-inflated}}_m = \delta_0(m)
\end{equation*}
is a single-point distribution concentrated at zero and the negative binomial part
\begin{equation*}
p^{\textrm{NB}}_m = \frac{(\lambda)_m}{m!}\left(\frac{h}{1+h}\right)^m\left(\frac{1}{1+h}\right)^{\lambda}
\end{equation*}
is a negative binomial distribution. The ZINB distribution is determined by three parameters with clear biological implications: the dropout rate $w$ which characterizes the proportion of the zero-inflated part due to both technical and biological effects, the mean burst size $h$ which describes the average number of transcripts synthesized in a single burst, and the maximum burst frequency $\lambda$ which represents the maximum number of occurrence of random bursts per mRNA lifetime. A more detailed discussion on the burst frequency will be given in the next section.

%From \eqref{distribution}, the steady-state probability that no mRNA is detected is given by
%\begin{equation}\label{zeroprob}
%p^{ss}_0 = w+(1-w)(1+h)^\lambda.
%\end{equation}

The ZINB distribution can exhibit three different types of shapes, as illustrated in Fig. \ref{distribution}. To clarify the conditions for the three types of shapes, we notice that the mode (maximum point) of the negative binomial part $p^{\textrm{NB}}_m$ is given by
\begin{equation*}
\mu_{\textrm{mode}} = \begin{cases}
0 & \textrm{when\;} \lambda<1, \\
[(\lambda-1)h] & \textrm{when\;} \lambda\geq 1,
\end{cases}
\end{equation*}
where $[x]$ denotes the integer part of $x$. In fact, the first type of shape occurs when $p^{ss}_0\leq p^{ss}_1$, that is,
\begin{equation*}
(\lambda-1)h \geq 1+\frac{w}{1-w}(1+h)^{\lambda+1}.
\end{equation*}
In this case, the dropout rate is small and the mode of the negative binomial part is large. Then the ZINB distribution peaks at the non-zero mode $[(\lambda-1)h]$ with no zero-inflation (Fig. \ref{distribution}(a)). The second type of shape occurs when $p^{ss}_0 > p^{ss}_1$ and $\mu_{\textrm{mode}}\leq 1$, that is,
\begin{equation*}
(\lambda-1)h < \min\{1+\frac{w}{1-w}(1+h)^{\lambda+1},2\}.
\end{equation*}
In this case, the dropout rate is large and the mode of the negative binomial part is small. Then the ZINB distribution peaks at zero with apparent or inapparent zero-inflation (Fig. \ref{distribution}(b)). The third type of shape occurs when $p^{ss}_0 > p^{ss}_1$ and $\mu_{\textrm{mode}}\geq 2$, that is,
\begin{equation*}
2 \leq (\lambda-1)h < 1+\frac{w}{1-w}(1+h)^{\lambda+1}.
\end{equation*}
In this case, both the dropout rate and the mode of the negative binomial part are large. Then the ZINB distribution becomes bimodal and peaks at both zero and the non-zero mode $[(\lambda-1)h]$ with apparent or inapparent zero-inflation (Fig. \ref{distribution}(c)).
\begin{figure}[!htb]
\centerline{\includegraphics[width=0.8\textwidth]{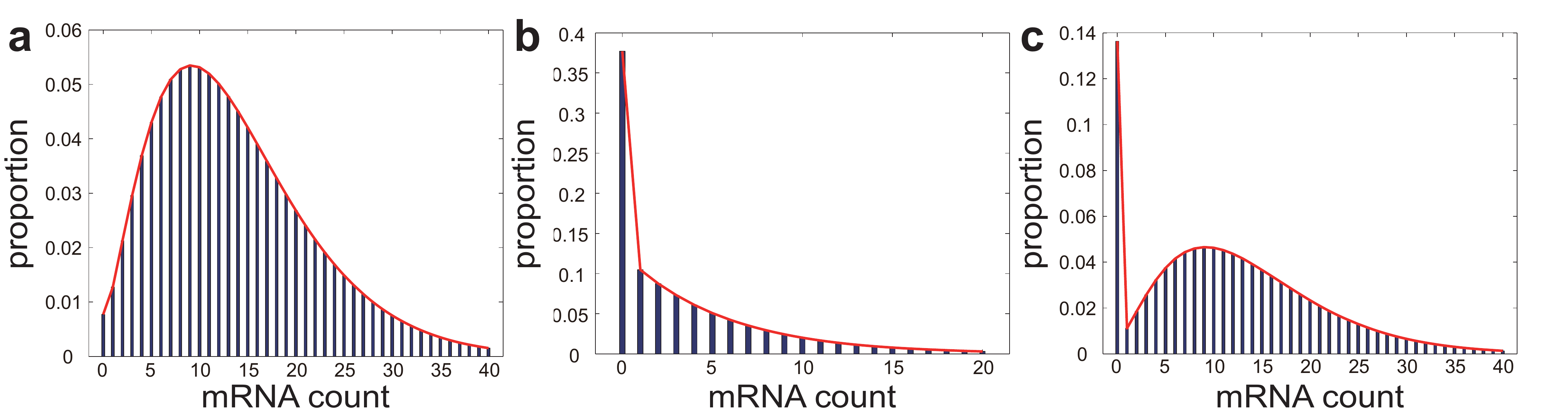}}
\caption{\textbf{Three different types of shapes for the ZINB distribution of mRNA abundance.} (a) The distribution peaks at a non-zero mode with no zero-inflation. (b) The distribution peaks at zero with apparent or inapparent zero-inflation. (c) The distribution exhibits bistability and peaks at both zero and a non-zero mode with apparent or inapparent zero-inflation.}\label{distribution}
\end{figure}

Three special cases should be paid special attention to. The first case occurs when the mean burst size $h\rightarrow 0$ and the maximum burst frequency $\lambda\rightarrow\infty$, while $\lambda h = \gamma$ is kept as a constant. In this case, we have
\begin{gather*}
(\lambda)_m\left(\frac{h}{1+h}\right)^m \rightarrow \gamma^m, \\
\left(\frac{1}{1+h}\right)^\lambda = \left(1-\frac{h}{1+h}\right)^\lambda \rightarrow e^{-\gamma}.
\end{gather*}
Then the ZINB distribution of mRNA abundance reduces to the zero-inflated Poisson (ZIP) distribution
\begin{equation*}
p^{ss}_m = w\delta_0(m)+(1-w)\frac{\gamma^m}{m!}e^{-\gamma}.
\end{equation*}
In fact, the ZIP distribution is also extensively applied in scRNA-seq data analysis \cite{chen2018umi} and its kinetic mechanism has been clarified in previous studies \cite{chong2014mechanism, klindziuk2018theoretical}. Our analytic theory shows that the ZIP model also naturally emerges from our three-state transcription model.

The second special case occurs when $b_2 = b_3 = 0$, which means that the switching from the active or inactive state to the dropout state is forbidden. In this case, the three-state model reduces to the classical two-state model without the dropout state \cite{paulsson2005models}. It is easy to verify that $\tilde a_1 = a_1$ and $w = 0$ in this regime. This shows that the dropout rate will vanish in the absence of the dropout state.

The last special case occurs when $a_3 = 0$, which means that the switching from the dropout state to the active state is forbidden. This is especially biologically relevant when the dropout state is understood to be the refractory (irreversibly silent) state found in recent single-cell experiments \cite{suter2011mammalian, bintu2016dynamics}. In this case, we also have $\tilde a_1 = a_1$ and thus the dropout rate is given by
\begin{equation*}
w = \frac{K_2}{K_2+(1+h)^{\lambda}},
\end{equation*}
where $K_2 = b_2/a_2$ is the equilibrium constant of gene switching between the inactive and dropout states. An increased equilibrium constant $K_2$ will result in a larger fraction of cells being in the dropout state and thus is expected to enhance the dropout rate $w$. Interestingly, our theory reveals a nontrivial quantitative relation between dropout events and transcriptional busting: an increased mean burst size $h$ or maximum burst frequency $\lambda$ will give rise to a decline in the dropout rate $w$. This relation provides novel insights into how and to what extent the burst size and burst frequency of the mRNA could reduce the dropout rate. The basic reason of such dependency is that an increase in the burst size and burst frequency will both promote rapid accumulation of mRNA from a low to a higher level, which is unfavorable to the occurrence of dropouts.

\section{Mean burst duration and burst frequency}
It has been shown that the mean burst size of the mRNA is given by $h = s/b_1$. Here we present a more detailed discussion on the burst frequency. In this section, we assume that the time-dependent mRNA abundance in an individual cell could be measured at a series of successive time points, and due to various technical factors, the mRNA expression is undetectable during some periods. Recall that each transcriptional burst is featured by a short transcriptionally active period followed by a long transcriptionally silent period. Mathematically, the mean burst duration, which is defined as the average time needed to complete a single burst (Fig. \ref{trajectory}(a)), can be computed as the inverse of the total probability flux between the active microstates and other (inactive and dropout) microstates \cite{jia2009general, jia2014allosteric}. From \eqref{distribution}, the total flux $J$ between the active microstates and other microstates is given by
\begin{equation*}
J = \left[\sum_{m=0}^\infty p^{ss}_{2,m}\right]a_1+p^{ss}_{3,0}a_3 = \alpha a_1,
\end{equation*}
where $0<\alpha\leq 1$ is a constant given by
\begin{equation*}
\alpha = \frac{\frac{a_1}{\tilde a_1}+\frac{a_3b_2}{\tilde a_1(a_2+a_3)}+(1+h)^{\lambda}-1}
{\frac{a_1}{\tilde a_1}+\frac{a_1b_2}{\tilde a_1(a_2+a_3)}+(1+h)^{\lambda}-1},
\end{equation*}
and thus the mean burst duration is given by
\begin{equation*}
\tau_{\textrm{burst}} = \frac{1}{J} = \frac{1}{\alpha a_1}.
\end{equation*}
Since the mRNA lifetime is the inverse of the mRNA degradation rate $v$, the mean burst frequency $\lambda_0$ of the mRNA, which is defined as the average number of occurrence of random bursts per mRNA lifetime, is given by the quotient of the mRNA lifetime $1/v$ and the mean burst duration $\tau_{\textrm{burst}}$:
\begin{equation*}
\lambda_0 = \frac{1}{v\tau_{\textrm{burst}}} = \frac{\alpha a_1}{v} = \alpha\lambda,
\end{equation*}
where $\lambda = a_1/v$ is the maximum burst frequency defined previously. Since $0<\alpha\leq 1$, the true mean burst frequency $\lambda_0$ is always smaller than the maximal burst frequency $\lambda$.

We next focus on three special cases. In the limiting case of $h\rightarrow 0$ and $\lambda\rightarrow\infty$, while $\lambda h = \gamma$ is kept as a constant, we have $\lambda_0\rightarrow\infty$. In this regime, random bursts occur very frequently but each burst only contributes a very small burst size. Due to large burst frequencies, the gene switches very rapidly between the active and inactive states, giving rise to a large number of ``futile" switches (Fig. \ref{trajectory}(b)).

In the special case of $b_2 = b_3 = 0$, the three-state model reduces to the classical two-state model without the dropout state \cite{paulsson2005models}. In this regime, we have $\alpha = 1$ and the mean burst frequency attains its maximum $\lambda_0 = \lambda$. In the presence of the dropout state, we have $b_2>0$ and $\alpha<1$. This shows that dropout events will lead to a reduction of the burst frequency by prolonging the transcriptionally silent periods.

The last special case occurs when $a_3 = 0$, which means that the switching from the dropout state to the active state is forbidden. In this case, we have $\tilde a_1 = a_1$ and
\begin{equation*}
\alpha = \frac{(1+h)^{\lambda}}{K_2+(1+h)^{\lambda}} = 1-w
\end{equation*}
is the proportion of the negative binomial part. Then the mean burst frequency is given by
\begin{equation*}
\lambda_0 = (1-w)\lambda.
\end{equation*}
This quantitative relation reveals how the dropout rate could affect the burst frequency.

\section{Over-dispersion of scRNA-seq data}
The simplest kinetic model of transcription is the classical birth-death process, which describes the synthesis and degradation of the mRNA. The steady-state distribution of the birth-death process turns out to be a Poisson distribution, whose mean and variance are equal. In bulk or single-cell RNA-seq experiments, read counts are always over-dispersed relative to Poisson: the variance is higher than the mean \cite{mccarthy2012differential, anders2010differential}.

In the literature, the dispersion, sometimes referred to as noise, in mRNA abundance within a cell population is often characterized by the Fano factor $\eta = \sigma^2/\langle m\rangle$, which is defined as the ratio of the variance $\sigma^2$ and the mean $\langle m\rangle$. A dispersion greater than one reveals a deviation from the Poisson distribution and thus serves as a characteristic signal of over-dispersion. Strictly speaking, the dispersion captures all sources of variation between samples, including contributions from technical factors leading to dropouts as well as real biological variation.

To calculate the mean and variance of mRNA abundance, we consider the generating function of the ZINB distribution:
\begin{equation*}
F(z) = \sum_{m=1}^\infty p^{ss}_mz^m = w+(1-w)\frac{q^\lambda}{(1-pz)^\lambda}.
\end{equation*}
Then the mean and variance can be recovered by taking the derivatives of the generating function:
\begin{equation}\label{meanvar}
\begin{split}
\langle m\rangle &= F'(1) = (1-w)\lambda h,\\
\sigma^2 &= F''(1)+F'(1)-F'(1)^2 = (1-w)[w\lambda^2h^2+\lambda h^2+\lambda h].
\end{split}
\end{equation}
Therefore, the dispersion in mRNA abundance is given by
\begin{equation*}
\eta = \frac{\sigma^2}{\langle m\rangle} = 1+h+w\lambda h,
\end{equation*}
where the constant term 1 is the dispersion of a Poisson distribution arising from individual births and deaths of the mRNA, the middle term $h$ describes the dispersion due to transcriptional burst sizes, and the last term $w\lambda h$ characterizes the dispersion due to the interaction between dropout events and transcriptional bursting. When there are no dropouts, the dispersion reduces to $\eta = 1+h$, which does not depend on the burst frequency \cite{paulsson2005models}. Interestingly, in the presence of dropout events, the dispersion positively depends on the three parameters: the dropout rate $w$, mean burst size $h$, and maximum burst frequency $\lambda$. This clearly reveals three different biophysical origins of over-dispersion.

Statistically, the three parameters involved in the ZINB distribution can be estimated in several different ways. The maximum likelihood estimation has been discussed in \cite{miao2018desingle}. Here we provide two additional approaches. In fact, the first three moments of the ZINB distribution can be recovered from the generating function as
\begin{equation*}
\begin{split}
\langle m\rangle\; &= F'(1) = (1-w)\lambda h, \\
\langle m^2\rangle &= F''(1)+F'(1) = (1-w)[\lambda(\lambda+1)h^2+\lambda h] \\
\langle m^3\rangle &= F'''(1)+3F''(1)+F'(1) \\
&= (1-w)[\lambda(\lambda+1)(\lambda+2)h^3+3\lambda(\lambda+1)h^2+\lambda h].
\end{split}
\end{equation*}
By analyzing scRNA-seq data, the first three moments of mRNA abundance can be estimated. Solving the above set of polynomial equations give the moment estimates of $w$, $h$, and $\lambda$.

When the mRNA levels across cells are relatively high, there is still another method to estimate the three parameters. From \eqref{distribution}, when $m\gg 1$, we have
\begin{equation*}
\frac{p^{ss}_{m+1}}{p^{ss}_m} = \frac{h}{h+1}\cdot\frac{m+\lambda}{m+1} \approx \frac{h}{h+1}.
\end{equation*}
This suggests that for any $k\geq 1$,
\begin{equation*}
p^{ss}_{m+k} \approx \left(\frac{h}{h+1}\right)^kp^{ss}_m.
\end{equation*}
Taking logarithm on both sides gives rise to
\begin{equation*}
\log p^{ss}_{m+k} \approx k\log\left(\frac{h}{h+1}\right)+\log p^{ss}_m,
\end{equation*}
which is a linear relation with respect to $k$. Therefore, we only need to calculate the logarithm of the steady-state probabilities at large mRNA copy numbers and then carry out a linear regression analysis with respect to the copy number difference $k$. The slope of the linear regression provides an estimate of the mean burst size $h$. Once $h$ is known, we can solve \eqref{meanvar} to obtain the estimates of the dropout rate $w$ and maximum burst frequency $\lambda$:
\begin{gather*}
w = \frac{\langle m\rangle}{\langle m\rangle+\eta-h-1}, \\
\lambda = \frac{(\eta-h-1)(\langle m\rangle+\eta-h-1)}{\langle m\rangle h}.
\end{gather*}
where $\eta = \sigma^2/\langle m\rangle$ is the dispersion.

\section{Discussion}\label{discussion}
In this work, we present a comprehensive analysis of a three-state transcription model with dropout events and over-dispersion based on the biochemical reaction kinetics underlying transcription. Using the multiscale simplification technique of decimation, we simplify the original Markovian model to a reduced one by removal of all fast states. It turns out that transcriptional bursts exactly correspond to the fast transition paths of the original model. Although the reduced model has a complicated topology, we obtain its steady-state analytic solution. The widely used ZINB or ZIP model of scRNA-seq data naturally emerges as the steady-state distribution of the reduced model. This provides a mesoscopic kinetic foundation of these statistical models. We further clarify the biological implications of the three parameters involved in the ZINB distribution: the dropout rate $w$, mean burst size $h$, and maximum burst frequency $\lambda$. In addition, we discover a nontrivial relation between dropout events and transcriptional bursting, which quantitatively reveals how and to what extent the burst size and burst frequency could reduce the dropout rate. Another relation reveals how dropout events could lower the burst frequency by prolonging the transcriptionally silent periods. The dispersion of scRNA-seq data is also investigated at the single-cell level and three different biophysical origins of over-dispersion are found. Finally, two statistical methods are given to estimate the three parameters involved in the ZINB distribution.

Our three-state transcription model is a minimal kinetic model that could account for the ZINB distribution of mRNA abundance. Recently, there has been some discussion on the role of various technical and biological effects on the apparent zero-inflation in scRNA-seq data \cite{chen2018umi, townes2019feature, svensson2019droplet}. In our minimal three-state model, zero-inflation is realized by the introduction of a dropout state, which may be either interpreted as an undetectable state due to technical factors or interpreted as a refractory state due to biological factors. In other words, our three-state model cannot distinguish whether zero-inflation is a consequence of technical or biological effects. If we would like to empirically decide between the two interpretations, a more realistic model that takes into account more complex features of stochastic transcription dynamics must be developed.

\begin{figure}[!htb]
\centerline{\includegraphics[width=0.65\textwidth]{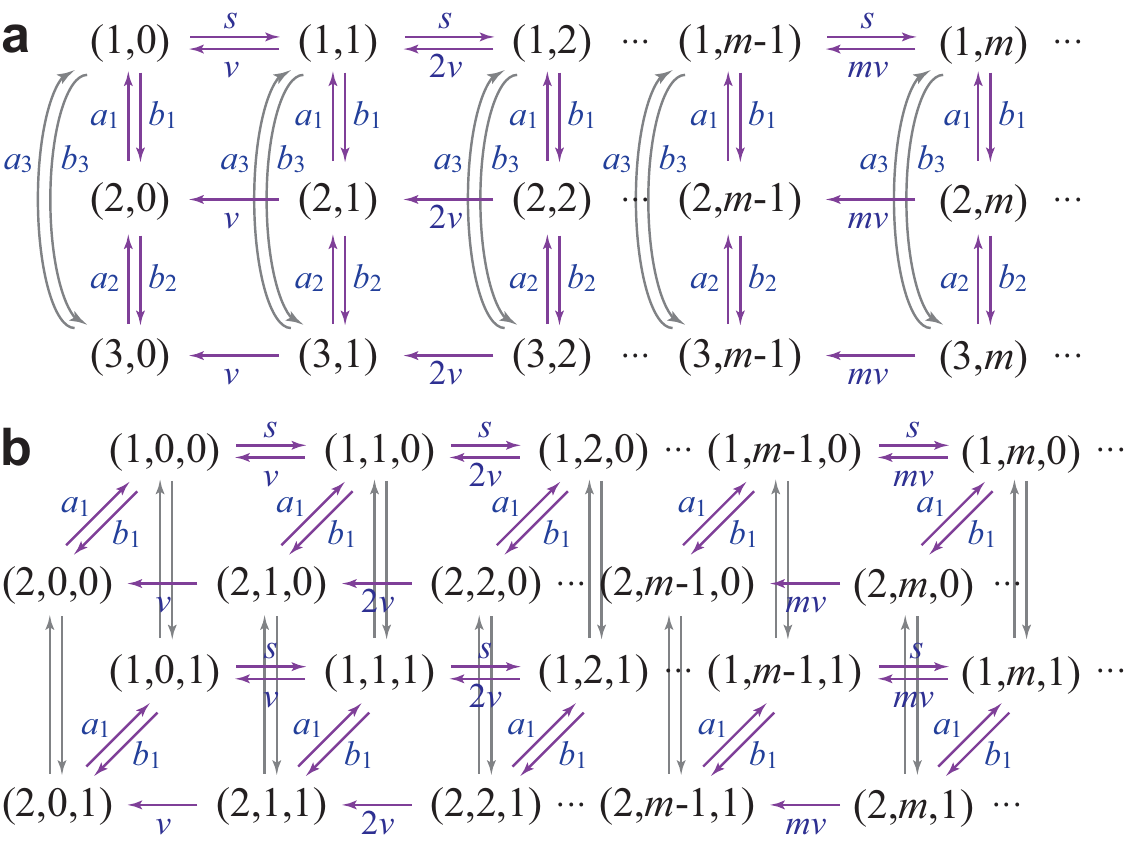}}
\caption{\textbf{More realistic models of transcription.} (a) A Markovian model of stochastic transcription involving gene switching among an active, an inactive, and a refractory state. The microstate of the gene of interest is described by an ordered pair $(i,m)$: the activity $i$ of the gene and the copy number $m$ of the mRNA. Here $i = 1,2,3$ correspond to the active, inactive, and refractory states, respectively. (b) A Markovian model of stochastic transcription with dropout events. The microstate of the gene of interest is described by an ordered triple $(i,m,k)$: the activity $i$ of the gene, the copy number $m$ of the mRNA, and the detection state $k$ of the transcriptional signal. Here $i = 1,2$ correspond to the active and inactive states, respectively, and $k = 1,0$ correspond to the detectable and undetectable states, respectively. }\label{threestate}
\end{figure}

If the dropout state is interpreted as a refractory state due to biological factors, then a more realistic model would be the Markovian model illustrated in Fig. \ref{threestate}(a), where microstates $(3,m)$, $m\geq 1$ are incorporated and transitions between microstates $(1,m)$, $(2,m)$, and $(3,m)$ are allowed. Here $(3,m)$ represents the microstate of having $m$ transcripts in an individual cell when the gene is in the refractory state. In fact, the minimal kinetic model depicted in Fig. \ref{model}(b) can be viewed as an approximation of the more realistic model when $a_2,a_3\ll v$. This can be understood intuitively as follows. Since $a_2,a_3\ll v$, the degradation of the mRNA is fast and the switching of the gene from the refractory state to the active or inactive state is slow. Once the gene is in the refractory state, before it could switch to the active or inactive state, the microstates $(3,m)$, $m\geq 0$ are already in rapid pre-equilibrium due to fast mRNA degradation and thus most of the probability is concentrated on microstate $(3,0)$.

If the dropout state is interpreted as an undetectable state due to technical factors, then a more realistic model would be the Markovian model illustrated in Fig. \ref{threestate}(b), where the microstate of the gene of interest is described by an ordered triple $(i,m,k)$: the activity $i$ of the gene with $i = 1,2$ corresponding to the active and inactive states, respectively, the copy number $m$ of the mRNA, and the detection state $k$ of the transcriptional signal with $k = 1,0$ corresponding to the detectable and undetectable states, respectively. In scRNA-seq experiments, the variable of interest is the copy number of detectable transcripts, which is given by
\begin{equation*}
N(i,m,k) = \begin{cases}
m, &\textrm{if}\;k = 1,\\
0, &\textrm{if}\;k = 0.
\end{cases}
\end{equation*}
This Markovian model allows transitions between detectable microstates $(i,m,1)$ and undetectable microstates $(i,m,0)$. Since dropouts are more frequent for cells with low mRNA expression levels \cite{kharchenko2014bayesian}, the transition rate from $(i,m,1)$ to $(i,m,0)$ should be a decreasing function of $m$ and the transition rate from $(i,m,0)$ to $(i,m,1)$ should be an increasing function of $m$. Within this framework, the minimal kinetic model depicted in Fig. \ref{model}(b) can be roughly viewed as an approximation of the more realistic model with all undetectable microstates $(i,m,0)$ combined as a single microstate $(3,0)$.

Besides the ZINB and ZIP models discussed in the present work, many other statistical models have also been proposed to analyze scRNA-seq data. Some commonly used models include but not limited to the Gaussian mixture model \cite{satija2015spatial}, Poisson-negative binomial mixture model \cite{kharchenko2014bayesian, fan2016characterizing}, Poisson-gamma mixture model \cite{huang2018saver}, Hurdle model \cite{finak2015mast}, zero-inflated log-normal model \cite{mcdavid2012data}, zero-inflated Gaussian mixture model \cite{paulson2013differential}, and Bayesian mixture model \cite{prabhakaran2016dirichlet, sun2019bayesian}. We anticipate that the mesoscopic kinetic mechanisms for these models could be clarified. A deeper understanding of the connection between the kinetic approach and the statistical approach is expected.

\section*{Acknowledgements}
The author acknowledges Michael Q. Zhang, Min Chen, and Cong Zhang at the University of Texas at Dallas, Yuxuan Liu at the University of Texas Southwestern Medical Center, Bochao Liu at Rutgers Cancer Institute of New Jersey, and Xuegong Zhang and Jiaqi Li at Tsinghua University for stimulating discussions. The author is also grateful to the anonymous reviewers for their valuable comments and suggestions which help the author greatly in improving the quality of this paper.

\section*{Appendix}
Here we provide the detailed derivation of the steady-state probability distribution of the reduced model depicted in Fig. \ref{decimation}(c). At the steady state, the steady-state probabilities of all microstates satisfy the following set of linear equations:
\begin{equation}\label{set}\left\{
\begin{split}
0 =&\; vp^{ss}_{2,1}+(a_2+a_3q)p^{ss}_{3,0}-(a_1p+b_2)p^{ss}_{2,0}, \\
0 =&\; b_2p^{ss}_{2,0}-(a_2+a_3)p^{ss}_{3,0}, \\
0 =&\; \sum_{k=0}^{m-1}a_1p^{m-k}qp^{ss}_{2,k}+(m+1)vp^{ss}_{2,m+1} \\
&\; +a_3p^mqp^{ss}_{3,0}-(a_1p+mv)p^{ss}_{2,m},\;\;\;m\geq 1. \\
\end{split}\right.
\end{equation}
By the second equation in \eqref{set}, we have
\begin{equation*}
(a_2+a_3)p^{ss}_{3,0} = b_2p^{ss}_{2,0}.
\end{equation*}
Inserting this equation into the first and third equations in \eqref{set} eliminates $p^{ss}_{3,0}$ and yields
\begin{equation}\label{setmod}\left\{
\begin{split}
0 &= vp^{ss}_{2,1}-\tilde a_1pp^{ss}_{2,0}, \\
0 &= \tilde a_1p^mqp^{ss}_{2,0}+\sum_{k=1}^{m-1}a_1p^{m-k}qp^{ss}_{2,k}+(m+1)vp^{ss}_{2,m+1}
-(a_1p+mv)p^{ss}_{2,m},\;\;\;m\geq 1,
\end{split}\right.
\end{equation}
where $\tilde a_1$ is the constant defined in \eqref{a1}. For convenience, set
\begin{equation*}
w_0 = \frac{\tilde a_1}{a_1}p^{ss}_{2,0},\;\;\;w_m = p^{ss}_{2,m},\;\;\;m\geq 1.
\end{equation*}
Then the two equations in \eqref{setmod} can be rewritten in a unified way as
\begin{equation}\label{master}
\sum_{k=0}^{m-1}a_1p^{m-k}qw_k+(m+1)vw_{m+1}-(a_1p+mv)w_m = 0,\;\;\;m\geq 0.
\end{equation}
To proceed, we introduce the generating function
\begin{equation*}
F(z) = \sum_{m=1}^\infty w_mz^m.
\end{equation*}
Then the algebraic equation \eqref{master} can be converted into the ordinary differential equation
\begin{equation*}
vF'(z) = \frac{a_1p}{1-pz}F(z),
\end{equation*}
whose solution is given by
\begin{equation*}
F(z) = A(1-pz)^{-a_1/v},
\end{equation*}
where $A$ is a constant. Therefore, $w_m$ can be recovered from the generating function $F$ as
\begin{equation*}
w_m = \frac{F^{(m)}(0)}{m!} = A\cdot\frac{p^m(a_1/v)_m}{m!}.
\end{equation*}
This shows that
\begin{equation*}
p^{ss}_{2,0} = A\cdot\frac{a_1}{\tilde a_1},\;\;\;p^{ss}_{2,m} = A\cdot\frac{p^m(a_1/v)_m}{m!},\;\;\;m\geq 1.
\end{equation*}

%%%%%%%%%% 参考文献格式 %%%%%%%%%%
\setlength{\bibsep}{5pt}
\small\bibliographystyle{nature}
%\bibliography{zeroinflated}

\end{document}